# STRUCTURE-DYNAMIC APPROACH IN NANOIONICS. MODELING OF ION TRANSPORT ON BLOCKING ELECTRODE

A.L. DESPOTULI,
*Institute of Microelectronics Technology, Russian Academy of Sciences,*
*142432, Chernogolovka, Moscow Region, Russia*
despot@ipmt-hpm.ac.ru

A.V. ANDREEVA
*Institute of Microelectronics Technology, Russian Academy of Sciences,*
*142432, Chernogolovka, Moscow Region, Russia*
andreeva@ipmt-hpm.ac.ru

Abstract. The work develops the structure-dynamic approach in nanoionics for detailed description of non-stationary ion-transport processes in irregular potential relief (direct problem) and interpretation of ionic properties and characteristics of nanosystems (inverse problem). Theoretical basis of the structure-dynamic approach, including structural model of interface space charge region, method of hidden variables and the criterion of smallness of external influence is presented. The approach is applied for the computational modeling of ion-transport in the region of ideally polarizable (blocking) solid electrolyte/electronic conductor (SE/EC) heterojunctions in ac and galvanostatic modes of external influence. In the computer experiments, the influence of (1) potential relief profile, (2) occupancy of mobile ion positions, and (3) current generator mode on ion-transport in the region of the SE/EC was investigated. Calculated data on the frequency behavior of impedance and capacitance of the SE/EC are interpreted in terms of mobile ion concentrations (hidden variables) on the crystallographic planes of SE in the interface space charge region. The effective thickness of the electric double layer (EDL) on the SE/EC is estimated. The generalized description of changes of charge distribution during the EDL relaxation in galvanostatic mode is given by the function of position of the center of mass of mobile ions. The simulated behaviors of processes as well as the ion transport characteristics of the SE/EC are in good agreement with known experimental data.

## 1. Introduction

Various physical-mathematical approaches to the problems of structure and properties of an electric double layer (EDL) can be found in the literature for systems with liquid electrolytes and ionic liquids [1-6]. The situation is different in nanoionics [7-9] which objects are solid state nanosystems and nanodevices with fast ionic transport (FIT). A radical change of mobile ions dynamics at the scale less than 1 nanometer was discovered in nanocrystal ionic conductors [10]. Specific dynamics of ion-transport processes at nano- and sub-nanometer scales can determine, for example, the frequency-capacitance characteristics of solid electrolyte (SE)/electronic conductor (EC) heterojunctions [11].

A generally used approach to the analysis of physical processes in FIT-devices consists in the solution of a differential equation system with constant values of diffusion ($D$) and drift ($\mu$) coefficients [12]. For example, this approach was used in modeling memristors [13,14], i.e. two-electrode nonlinear nanodevices with memory [15]. The constancy of $D$ and $\mu$ values assumes that the depth ($\eta$) of a crystal potential relief in which ions move remains invariable at distances ($l$) considerably exceeding the length ($\Delta$) of ion elementary jump (regular structures). However, this condition can be violated in the region of heteroboundaries.

The central challenge in nanoionics of advanced superionic conductors (AdSICs) [8], namely the influence of atomic structure on the ion-transport and polarization processes in the space charge region of a heterojunction, has not been properly considered in terms of the EDL-theory. The AdSIC crystal structures are close to optimum for FIT [8]. This defines the record-high ion-transport characteristics such as ionic conductivity $\approx$ 0.3 Ohm$^{-1}$ cm$^{-1}$ (300 K) and activation energy of ionic transport about $4k_BT_{300} \approx$ 0.1 eV. For example, in AdSICs of the $\alpha$-RbAg$_4$I$_5$-family, the mobile cations Ag$^+$ with total concentration $\approx 10^{22}$ cm$^{-3}$ move within the crystallographic tunnels formed by the "rigid" I$^-$ anion sub-lattice, overcoming relatively low potential barriers $\eta_v \approx$ 0.1 eV. One dimensional (1D) transport tunnels for mobile ions exist in biological objects, synthetic organic membranes and some monocrystals [16]. The classification diagram in the "ionic conductivity - electronic conductivity" coordinates [8,17] shows that ionic conductors of the AdSIC-class intersect with the class of «solid electrolytes» (SE) which are defined as ionic conductors-electronic dielectrics. On the basis of SE with "low" ion-transport characteristics (regular crystal potential relief with $\eta_v \sim$ 0.5 eV) nanocomposites with enhanced integral ionic conductivity were created [18]. However, their conductivity is considerably lower than in the AdSIC bulk.

Interphase boundaries influence the transport properties of SE in a complicated manner [19-21]. The appearance of surface ionic conductivity caused by a reduction in the energy of point defects formation at the crystal/vacuum heteroboundaries was predicted by K. Lehovec [22]. The opposite effect, namely conductivity decrease, should dominate in AdSIC nanosystems, where the bulk AdSIC structure (close to optimal for FIT) is disturbed on the disordered interfaces [8,11]. A sharp increase in the activation energy of ionic conductivity was found in thin ($\leq$ 40 nm) non-epitaxial films of the $\alpha$-RbAg$_4$I$_5$ AdSIC [23].

Residual stresses on heteroboundaries change the parameters of the crystal structure [24]. The balance between the interfacial energy and bulk strain energy near the interface governs the transition from coherent to semicoherent and incoherent heteroboundaries [25]. The gradients of stresses and elastic deformations near interfaces [26,27] should distort the potential relief of AdSIC, creating an irregular relief in the $x$ direction perpendicular to the heteroboundary. In AdSIC the potential barriers heights $\eta_v$ strongly depend on the relation between the radius of a mobile ion and the cross-section size of the FIT tunnel determined by interatomic distances in this region [28]. Because linear distortions of an elementary cell during deformations can be about 1 % (about 0.005 nm for AdSIC $\alpha$-AgI [29]), then, in view of estimations for variation of $\eta$ [28], these values can reach about 1 eV nm$^{-1}$ near AdSIC/EC heteroboundaries.

The information on the ion-transport processes in the space charge region at the AdSIC-SE/EC heterojunctions (cause) and ion-transport properties and characteristics of nanosystems (consequence) can be obtained from experimental data with the help of the FIT-theory by solving direct (cause $\rightarrow$ consequence) and inverse (consequence $\rightarrow$ cause) problems. However, if an adequate physical model is not developed, the formalization of ionic transport description in the EDL, e.g. with the help of equivalent electric circuits in impedance spectroscopy, can provide the systematization of data only for a limited range of frequencies.

A structure-dynamic approach [30,31] proposed recently for detailed description of ion-transport processes in irregular potential relief (direct problem) and interpretation of ion-transport processes and characteristics of nanosystems (inverse problem) includes: (i) a structural model that interconnects the relaxation rate of an EDL and ion movement in a crystal potential relief of a "rigid" sub-lattice of SE distorted at the SE/EC; (ii) choice of "hidden" variables, providing the description of ion-transport processes and solution of the inverse problems in terms of mobile ion concentrations on the crystallographic planes of SE in the space charge region of a heterojunction; (iii) physico-mathematical formalism that operates with "hidden" variables, basing on the concept of a detailed balance and a kinetic equation in the form of the particle conservation law.

In this work the structure-dynamic approach [30,31] is applied for the investigation (in the computer experiments) the influence on the ion-transport processes and characteristics of such factors as 1) the parameters of exponential potential profile in the region of the EC/SE-AdSIC ideally polarizable (blocking) heterojunctions, 2) the occupancy degree of the crystallographic positions accessible for mobile ions, and 3) parameters of mode of external generator. The obtained results are compared with the known experimental data.

## 2. Theoretical Basis of the Structure-Dynamic Approach

In section 2 an initial basis of the theory of ionic transport in an irregular potential relief is formulated for the EC/SE-AdSIC blocking heterojunctions which are considered as nanosystems. We state the problem and deduce a system of differential equations for designated dynamic nanosystems. The problem is formalized as follows.

### 2.1. *Structural model and method of hidden variables*

Consider an EC/AdSIC heterojunction as the simplest layered nanostructure in which the potential relief in AdSIC represents a sequence of potential barriers of different heights $\eta$. The number of layers in the heterojunction equals $M$. The irregular potential relief $\eta(x)$ extends from EC along the $x$-coordinate for several nanometers, which is determined by the influence of interface atomic structure. The coordinate $x_0 = 0$ corresponds to the surface of EC blocking electrode and $x = x_M$ corresponds to the EDL right edge. The barrier minima $\eta(x_j) = 0$ are located in the points $x_j$ ($j = 1,2…M$), and the maxima are in ($x_j - \Delta/2$) along the $x$-coordinate, where $\Delta = x_{j+1} - x_j$. The $\eta$ value decreases with the distance from $x_0$, and at $x = x_M - \Delta/2$ it becomes equal to the barrier height $\eta_v$ in the AdSIC bulk. Along the $y$ and $z$-coordinates the quantity $\eta$ ($x_j - \Delta/2, y_p, z_q$), like along the $x$-coordinate, is a discrete function ($p$ and $q$ are integer numbers from 1 to $\infty$) determined by the AdSIC periodicity. The values of $\eta$ ($x_j - \Delta/2, y_p, z_q$) = $\eta$ ($x_j - \Delta/2$) = $const$ for any $p$ and $q$, i.e. layers are macroscopic along the $y$- and $z$-coordinates.

Now, apply a structural model for the analysis of ion-transport processes at the EC/SE-AdSIC blocking coherent (structure-ordered) heterojunction (Fig. 1). Designate the heterojunction as EC/{$X^j$} ($j =1,2…M$), where {$X^j$} is a sequence of crystallographic planes in the interface region with a distorted potential relief (at $j < M$) relative to the bulk. In the EC/{$X^j$} ($j =1,2…M$) region, the barrier heights $\eta_{j,j+1}$ decrease with increasing index $j$.

The interface structural coherence means that the {$X^j$} ($j =1,2…M$) crystallographic planes within the "rigid" anionic sub-lattice of SE-AdSIC (adjacent to the EC) are orientated parallel to the EC crystallographic surface plane $X^0$ and the normal of the planes coincides with the orientation of fast ion transport tunnels in AdSIC [32]. Atomic sharp coherent EC/SE-AdSIC blocking heterojunctions can be formed by epitaxial growth methods. The structural model of a heterojunction partitions the 3D space of an ionic conductor (adjacent to EC) into a number of planar layers, just as the layered lattice gas model in the equilibrium EDL theory [1-3]. A distinguishing feature of the structure model is the presence of potential barrier maxima of $\eta$ ($x_j - \Delta/2, y_p, z_q$) inside each layer and $\eta_{j,j+1}$ decreasing with the $j$ rising [30,31].

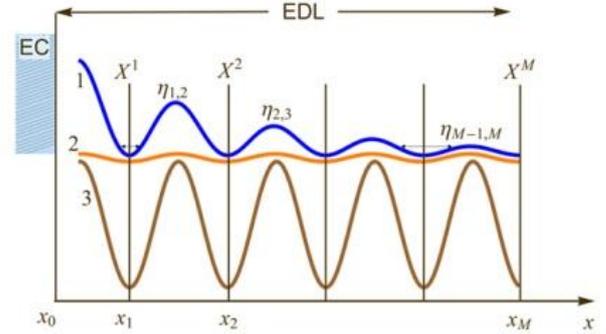

Fig.1. Schematic presentation of the crystal potential relief for EC/{$X^j$} heterojunction (1), AdSIC bulk (2), and classical ionic conductor (3).

In the Cartesian $xyz$-system the cation state $j \in S$ ($S$ is a linearly ordered set of states) is defined uniquely by the coordinate $x_j \in X^j$ (Fig. 1). Assume that the cation chemical potential does not depend on $x_j$, therefore without an external influence (a current generator) on the $X^0$ and $X^M$ edge planes of heterojunction, the equilibrium cation concentrations on all planes {$X^j$} ($j = 1,2…M$) are taken equal to $n_0$. The cations jump between the adjacent planes $X^j$ and $X^{j+1}$ over a potential barrier of height $\eta_{j,j+1}$. The potential reliefs in a classical ionic crystal ($\eta_v > 1$ eV), in the AdSIC bulk ($\eta_v < 0.2$ eV) and on the EC/SE-AdSIC blocking heterojunction (irregular potential relief) are compared in Fig. 1.

The proposed structural model reflects the basic features of structure and ionic transport in AdSICs, e.g. $α$-AgI and $α$-RbAg$_4$I$_5$ with a set of vacant positions on the crystallographic planes $\{X^j\}$ ($j$ =1,2…$M$) accessible for mobile silver cations in the "rigid" iodine sub-lattice (Fig. 2).

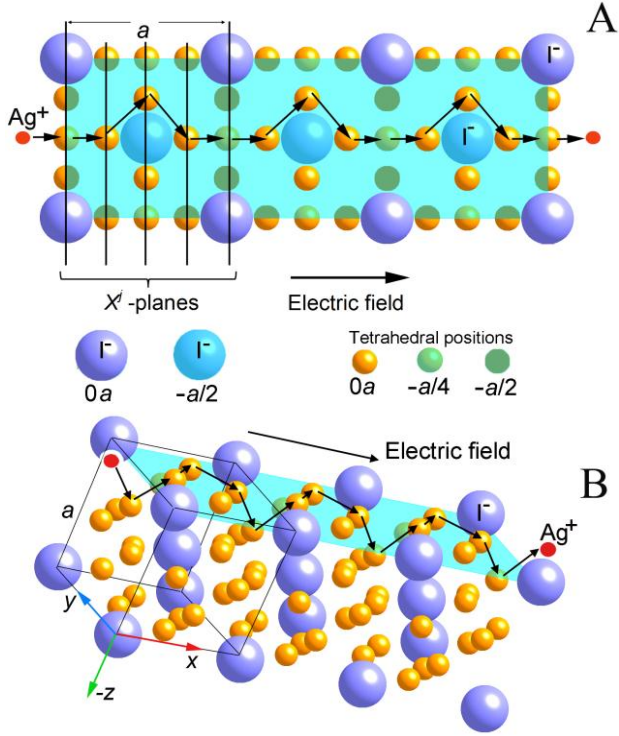

Fig.2. The structure of $α$-AgI AdSIC (Im3m, $a$ = 0.509 nm) with a rigid iodine sub-lattice and tetrahedral positions accessible for movement of Ag$^+$-cations. One of possible ways of Ag$^+$-cations movement along the direction of an external electric field is schematically shown by arrows in the projection <001> (Fig.2A) and in the crystal bulk (Fig.2B). The tetrahedral positions are on the $X^j$ crystallographic planes. Different colors mark different $z$-coordinates of I-ions and tetrahedral vacant positions relatively to the drawing plane.

In the AdSICs the cation movement between neighboring $X^j$ and $X^{j+1}$ planes occurs over low potential barriers. A set of routes for cation transport, e.g. in $α$-AgI through vacant tetrahedral positions [33], forms a system of zigzagging FIT-"tunnels". The projections of cation movement occur along the general <100> direction, normal to the $X^j$ planes. In the $α$-AgI (Im3m, $a$ = 0.509 nm) the distance $Δ$ between the adjacent parallel planes $X^j$ and $X^{j+1}$ is ≈0.127 nm. The unit cell contains two Ag$^+$ cations per 12 tetrahedral positions (Fig. 2), therefore the total concentration of tetrahedral crystallographic positions accessible for Ag$^+$ on the $X^j$ <100> planes is $n^* ≈ 6n_0$, where the concentration of silver ions is $n_0 ≈ 2·10^{18}$ m$^{-2}$ on the $X^j$ planes [34]. The ratio $n^*/n_0$ differs for different AdSICs. For example [35], $α$-RbAg$_4$I$_5$ with a record high ionic conductivity at 300 K has $n^*/n_0 ≈$ 4.

In the structural model the concentrations of mobile ions in the minima of potential relief $x_j \in X^j$ are considered as real but unobserved, i.e. hidden, variables. They are such real physical quantities which the modern experimental techniques cannot measure yet. The method of hidden variables is used in a number of scientific disciplines [36], for example, at the reconstructing equations of motion from experimental data for systems where not all necessary variables have been observed [37,38].

The $η$ distribution in the EC/$\{X^j\}$ region is determined by the modeling dependence

$$η_{j,j+1} = a_0 + b_0 \exp(-j/c), \quad j =1,2…20 \quad (1)$$

with parameters $a_0$ = 0.3 eV, $b_0$ = 0.4 eV and $c$ = 4.5. Thus, the model heterojunction is a nanosystem with twenty one $X^j$ planes. Expression (1) corresponds to crystallographic narrow interfaces for which stresses and appropriate deformations are distributed for a distance of several lattice parameters from the interface [27,39]. The height difference of potential barriers of 0.7 - 0.3 eV on the heteroboundary corresponds to the transition from SE with low ion-transport characteristics to superionic conductors [17] with inferior characteristics compared to AdSICs. The choice of (1) is justified because characteristic frequencies of jumps of cations in the AdSICs bulk are 10$^9$ - 10$^{10}$ Hz (300 K) whereas the laboratory samples of supercapacitors based on the AdSICs [11,40] have the best frequency-capacitance characteristics at the frequencies ~ 10$^5$ Hz (300 K), which corresponds to $η_{1,2} > 0.35$ eV.

## 2.2. *Physico-mathematical formalism*

The motion of classical particles performing a random walk in the lattice (stochastic processes over discrete states [41]) can be described by the kinetic equation and detailed-balance relation [42,43]

$$dP_i/dt = \sum (P_j w_{j \to i} - P_i w_{i \to j}); i,j \in S \quad (2)$$

with summation on all states $j \neq i$, and

$$w_{i \to j}/w_{j \to i} = \exp[(E_i - E_j)/k_B T] \quad (3)$$

where values $w_{i \to j} dt$ are a set of probabilities of transitions between the states $i \to j$ in the time interval $dt$. The transitions $i \to j$ should be regarded as instantaneous events [41]. For an EC/{$X^j$} heterojunction only transitions between the adjacent planes $X^j \leftrightarrow X^{j+1}$ over the barriers $\eta_{j,j+1}$ are possible (linear system). Therefore, $P_j$ is the probability to find a cation in $X^j$, and ($E_i - E_j$) is the cation energy difference for adjacent states, $k_B$ is the Boltsman constant and $T$ is temperature (in K).

The principle of detailed balance is valid for the Markov stochastic processes which statistical properties in the time moment $t + dt$ ($dt > 0$) depend only on the processes at $t$. For the Markov processes in the EC/{$X^j$} ($j = 1,2…M$) nanosystem the condition $P_j(t) \propto n_j(t)$ is satisfied because $P_j(t) = n_j(t)/\sum n_i(t)$ with summation over all $i = 1,2…M$, where $n_j(t)$ are time-dependent concentrations of mobile cations in the potential relief minima $x_j$ ($x_j \in X^j$). The $\sum n_i(t)$ is determined by the current generator affecting the nanosystem.

For the EC/{$X^j$} ($j = 1,2…21$) linear nanosystem with the distribution (1), the relation (2) can be written as the conservation law of mobile cations

$$dn_j/dt = - n_j w_{j \to j+1} - n_j w_{j \to j-1}$$
$$+ n_{j-1} w_{j-1 \to j} + n_{j+1} w_{j+1 \to j} \quad (4)$$

where $j = 1,2…20$. For the right edge of the nanostructure ($j = 21$), the boundary condition is

$$dn_{21}/dt = - n_{21} w_{21 \to 20}$$
$$+ n_{20} w_{20 \to 21} - I_{21}(t)/en_0 \quad (5)$$

where $e$ is the absolute value of an electron charge, and $I_{21}(t)$ is the density of the cation current created by a current generator on the $X^{21}$ plane. The probabilities of transitions in unit time $w_{0 \to 1} = 0$ and $w_{1 \to 0} = 0$ (boundary condition for EC blocking electrode). The minus in (5) shows that at $dn_{21}/dt < 0$ (increasing of cations deficit on $X^{21}$) the vector of cation current $I_{21}(t)$ and a positive direction of the $x$-axis (+$x$) coincide. Along with ion transitions on crystallographic positions between adjacent parallel $X^j \leftrightarrow X^{j+1}$ planes, ions can move also through crystallographic positions of the same $X^j$ plane [44,45]. Such transitions give the contributions into macroscopic potentials on the $X^j$ planes. It corresponds to the mean field approximation which is used in the majority of the EDL-theories [1-6].

The probabilities $w$ in (4) and (5) depend on the direction of an electric field. If the field direction coincides with +$x$, then cation jumps $j \leftrightarrow j+1$ occur from the minimum $x_j$ of a potential relief to the minimum $x_{j+1}$ which is displaced downwards by the value $e^2 \Delta \sum (n_k - n_0)/\varepsilon_0 \varepsilon_{j,j+1}$ relative to $x_j$ ($j+1 \leq k \leq M$). Therefore, the barrier height for transitions $j \to j+1$ ($j+1 \to j$) is less (more) than $\eta_{j,j+1}$. The field additive $\Omega_{j+1,M}(t)$ to $\eta_{j,j+1}$ depends on $t$ through a set of hidden variables $n_k(t)$:

$$\Omega_{j+1,M}(t) \equiv e^2 \Delta \sum (n_k - n_0)/2\varepsilon_0 \varepsilon_{j,j+1}, \quad (6)$$

where ($j+1 \leq k \leq M$, $\varepsilon_0$ is the electric constant and $\varepsilon_{j,j+1}$ is the effective dielectric susceptibility of the layer between $X^j$ and $X^{j+1}$ planes.

The values $\varepsilon_{j,j+1}$ are determined by cation displacement in the potential relief minima at the electric field strength of a unitary value. Therefore, dielectric polarization should increase with a reduction of the potential relief depth $\eta$, which is schematically shown by different arrow lengths for the potential relief minima with the coordinates $x_1$ and $x_{M-1}$ (Fig. 1). In classical ionic crystals, value $\eta$ is several eV and the relative dielectric permeability $\varepsilon \approx 5$. In the AdSIC bulk $\eta \approx 0.1$ eV and $\varepsilon \approx 90$ [46], which can be connected with large polarization displacements of cations in potential wells of small depth. It can be assumed that for an EC/{$X^j$} with an irregular potential relief $\varepsilon_{j,j+1} \propto 1/\eta_{j,j+1}$ [11,47].

The probabilities $w_{j \to j+1}$ in (3) are proportional to frequencies $v_{j \to j+1}$ of the $j \to j+1$ transitions, and $v_{j \to j+1}$ are proportional to the maximum frequency of crystal lattice oscillations, that is about Debye frequency $v_D \approx 10^{12}$ s$^{-1}$ (300 K) [48]:

$$w_{j \to j+1}(t) \propto v_{j \to j+1}$$

$$v_{j \to j+1} \approx v_D \exp[(-\eta_{j,j+1} + \Omega_{j+1,M})/k_B T] \quad (7)$$

$$v_{j+1 \to j} \approx v_D \exp[(-\eta_{j,j+1} - \Omega_{j+1,M})/k_B T]. \quad (8)$$

Relations (7) and (8) introduce dynamics into kinetic equations (4) and (5) through the dependence $\Omega$ on $n_j(t)$. In (7) and (8) the probabilities $w \propto v$, that implies the frequencies of cation transitions $v$ do not depend on the occupancy of adjacent positions (n* >> $n_0$ approximation).

Let take into account the influence of occupancy degree of the neighboring positions on $w$. In $\alpha$-AgI the occupancy degree of tetrahedral positions for Ag$^+$ mobile cations is about 1/6 [33], then the total number of positions for Ag$^+$ is equal to 6 $n_0$ on each $X^j$ plane (Fig. 2). Therefore we consider that

$$w_{j \to j+1}(t) \propto f_{j \to j+1}(t)\, \rho_{j+1}(t) \quad (9)$$

where $f_{j \to j+1}$ is the frequency with which cations reach the potential barrier top and $\rho_{j+1} = (6n_0 - n_{j+1})/6n_0$ designates the probability of a vacant position on the $X^{j+1}$ plane. Taking (9) into account, expressions (7) and (8) can be written as

$$v_{j \to j+1} \approx v_D\, \rho_{j+1} \exp[(-\eta_{j,j+1} + \Omega_{j+1,M})/ k_B T] \quad (10)$$

$$v_{j+1 \to j} \approx v_D\, \rho_j \exp[(-\eta_{j,j+1} - \Omega_{j+1,M})/ k_B T]. \quad (11)$$

Therefore (4) can be written as

$$dn_j/dt = e(- n_j f_{j \to j+1}\rho_{j+1} - n_j f_{j \to j-1}\rho_{j-1}$$
$$+ n_{j-1} f_{j-1 \to j}\rho_j + n_{j+1} f_{j+1 \to j}\rho_j). \quad (12)$$

The terms in (12) represent the current density with a sign defined by "from $j$ minimum" (sign minus) or "into $j$ minimum" (sign plus). Further in the text we shall use definitions of the current densities relatively to the $x$ axis direction. Namely, $en_j f_{j \to j+1}\rho_{j+1} \equiv I_{j \to j+1} > 0$ as the current density over the barrier $\eta_{j,j+1}$ in the positive direction of the $x$ axis, i.e. +$x$. The quantities $en_{j+1} f_{j+1 \to j}\rho_j \equiv I_{j \leftarrow j+1} < 0$ are the current densities over the barrier $\eta_{j,j+1}$ in the negative direction of the $x$ axis, i.e. -$x$. The net densities of cation currents over potential barriers $\eta_{j,j+1}$ separating the $X^j$ and $X^{j+1}$ planes are $I_{j \to j+1} + I_{j \leftarrow j+1} \equiv I_{j,j+1}$. In much the same way, we can write $en_{j-1} f_{j-1 \to j}\rho_j \equiv I_{j-1 \to j} > 0$, $en_j f_{j \to j-1}\rho_{j-1} \equiv I_{j-1 \leftarrow j} < 0$ and $I_{j-1 \to j} + I_{j-1 \leftarrow j} \equiv I_{j-1,j}$. Then, expression (12) can be rewritten as

$$dy_j/dt = - (I_{j \to j+1} + I_{j \leftarrow j+1})$$
$$+ (I_{j-1 \to j} + I_{j-1 \leftarrow j}) = - I_{j,j+1} + I_{j-1,j}, \quad (13)$$

where $y_j \equiv (n_j - n_0)/n_0$ are dimensionless variables (relative change of cation concentrations on $X^j$. The minus in (13) means that $I_{j,j+1}$ reduces the concentration of cations on $X^j$ when the vector of cation current $I_{j,j+1}$ and +$x$ coincide. For the layer between $X^j$ and $X^{j+1}$, the electric field strength $F_{j+1,j}$ satisfies the Gauss law:

$$F_{j+1,j}(t) = - (en_0 \Sigma y_k)/2\varepsilon_0\varepsilon_{j,j+1}, \quad (14)$$

where $j+1 \leq k \leq M$. The minus in (14) shows that at $y_k < 0$ (deficit of cations on $X^j$ planes) the vector directions of $F_{j+1,j}$ and +$x$ coincide (Fig.1).

## 2.3. *The criterion of smallness of external influence*

The principle of detailed balance is strictly satisfied in equilibrium states of systems. According to [49], a measure of deviation from an equilibrium state can be introduced for stationary conditions where the principle of detailed balance is violated. Charge and voltage on the blocking heterojunction depend on the current density integral $\int I(t)dt$, therefore the criterion of smallness of external influence on current densities is chosen as a measure of small deviation from the system equilibrium state in present work.

During the external influence and after its termination a non-equilibrium distribution of the space ionic charge $\{n_j(t)\}$ relaxes to an equilibrium one $\{n_j(\infty)\}$. If the condition of charge conservation holds on the EC/$\{X^j\}$ ($j =1,2…M$) blocking heterojunction

$$\Sigma n_j(t) = \Sigma n_j(\infty)\quad 1 \leq j \leq M. \quad (15)$$

then relaxation leads to the same equilibrium distribution $\{n_j(\infty)\}$ which results from the flow of two opposite-directed currents $I_{j \to j+1}$ and $I_{j \leftarrow j+1}$ over each $\eta_{j,j+1}$ barrier. The resulting currents $I_{j,j+1} \equiv I_{j \to j+1} + I_{j \leftarrow j+1}$ become equal to zero at $\{n_j(\infty)\}$ where the principle of detailed balance $|I_{j \to j+1}| = |I_{j \leftarrow j+1}|$ is rigorously fulfilled. For the distribution $\{n_j(\infty)\}$ the modulus $|I_{j \to j+1}|$ can be estimated [39]

$$|I_{j \to j+1}| \approx en_0 \exp(-\eta_{j,j+1}/k_B T) \equiv I_{ex} \quad (16)$$

if the transitions $j \to j \pm 1$ are instantaneous events. The characteristic quantity $I_{ex}$ can be named an exchange current density on the potential barrier $\eta_{j,j+1}$. For example, $I_{ex}$ is of the order of $2.7 \cdot 10^{-1}$, $6.3 \cdot 10^2$ and $1.5 \cdot 10^6$ A m$^{-2}$ for barriers $\eta_{j,j+1}$ with the heights 0.7, 0.5 and 0.3 eV, respectively ($n_0 = 10^{18}$ m$^{-2}$, $T = 300$ K and $v_D = 10^{12}$ s$^{-1}$). The criterion of smallness of external influence which brings a heterojunction to a non-equilibrium state can be expressed through ionic currents on barriers $\eta_{j,j+1}$.

$$|I_{j,j+1}|/|I_{j\to j+1}| \ll 1, \quad j = 1,2\ldots 20 \quad (17)$$

and the criterion of smallness of current $I(t)$ of an external current generator

$$|I(t)|/|I_{j\to j+1}| \ll 1, \quad j = 1,2\ldots 20 \quad (18)$$

The fulfillment of (17) and (18) inequalities means that deviations from an equilibrium state are insignificant; therefore, the principle of detailed balance can be used. Inequalities (17) and (18) correspond to the results reported in [49] where the norm of transport matrix with probability currents as elements was taken as a measure of detailed balance violation in the stationary non-equilibrium states. In case of EC/{$X^j$} heterojunctions the current densities $I_{j,j+1}(t)$ are elements of a transport matrix.

## 3. Computer Simulation and Discussion of Results

In this section, the results of numerical modeling (on the basis of the structure-dynamic approach) of ion-transport processes, properties and characteristics (capacitance, impedance) of EC/SE heterotjunctions are presented. The obtained results are compared with the known experimental data and interpreted in terms of the "hidden" variables.

### 3.1. *The basic properties of solutions of ion-transport equations*

The numerical solution of the system of equations (4) - (8) with boundary conditions, which specify the external influence on EC/{$X^j$}, gives a set of time-dependent hidden variables {$n_j(t)$} which completely define the state of a model nanosystem. Some combinations of $n_j(t)$ correspond to observable physical quantities, and others - to the hidden quantities. For example, impedance $Z$ of a heterojunction is an observable quantity, while the phase shift between the external current and potential of the $X^j$ plane in EDL is a hidden quantity.

In conditions of charge conservation (15) and termination of external influence (the heterojunction is disconnected from the load) any non-equilibrium distribution {$n_j(t)$} should relax at $t \to \infty$ to the same equilibrium distribution {$n_j(\infty)$}. It was confirmed by the calculations, e.g. in the $n^* \gg n_0$ approximation (Fig. 3). Fig. 3 shows a change in the $|y_j - Y_j|$ differences where the time-dependent dimensionless state functions $y_j \equiv (n_j - n_0)/n_0$ and $Y_j \equiv (n_j - n_0)/n_0$ belong to different sets of the {$y_j(t)$} and {$Y_j(t)$} and correspond to two strongly differing initial non-equilibrium distributions of ionic space charge. It is seen that for the {$X^j$} ($j$ =1,2…21) heterojunction, where the highest potential barrier is 0.7 eV, the relaxation time $\tau_r$ (non-equilibrium space charge) is $\tau_r < 1$ s at $T = 300$ K and $n_0 = 10^{18}$ m$^{-2}$.

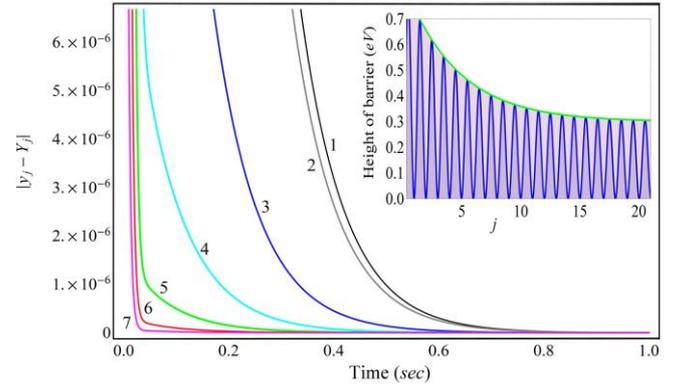

Fig.3. Time dependence of the modulus $|y_j - Y_j|$ ($j \leq 7$) of the EC/{$X^j$} ($T = 300$ K, $n_0 = 10^{18}$ m$^{-2}$ at space charge relaxation for two initial non-equilibrium distributions of cations. The sequence of potential barriers (insert) is defined by (1). The state functions {$Y_j(t)$} are calculated for initial distribution $n_2 = 0.999\, n_0$, and the rest $n_j(0) = n_0$ at $t = 0$. The functions {$y_j(t)$} are calculated for initial distribution $n_{21} = 0.999\, n_0$, and the rest $n_j(0) = n_0$ at $t = 0$. The graph number corresponds to the $j$ index.

Changes of moduli $|y_j(t)|$ at the space charge relaxation on the EC/{$X^j$}heterojunction are shown in log-log scale in Fig. 4. The initial distribution of cations {$n_j(t)$}in EDL is specified by a small non-equilibrium concentration of the charge on the edge plane $X^{21}$, i.e. $n_{21}(0) < n_0$, and on other planes $n_j(0) = n_0$ ($j \neq 21$). For all potential relief minima $x_j$ ($j = 1,2 \ldots 21$), $|y_j(t)|$ are calculated by expressions (4), (5), (10) and (11) which take into account the dependence of probabilities $w$ of the $j \leftrightarrow j \pm 1$ transitions on the degree of filling of the nearest minima ($n^* = 6n_0$ approximation). For comparison, some values $|y_j(t)|$ ($j$ =1, 2, 3 and 21) are also calculated by the (4) - (8) formulas, i.e. without taking the degree of $X^j$ filling into account ($n^* \gg n_0$ approximation).

The analysis of the data (Fig. 4) shows that (i) at $t \approx 0$, changes in $y_j(t)$ with small indices, e.g. $j$ =1, 2, 3 and 4 (large $\eta_{j,j+1}$), occur without lag relatively changing of $y_j(t)$ with large indices, e.g. $j = 20$ and 21 (small $\eta_{j,j+1}$). This means that the long-range electric field created by charge of the $X^{21}$ edge plane determines EDL-relaxation at $t \approx 0$; (ii) at the $\{n_j(t)\} \to \{n_j(\infty)\}$ relaxation, changes in $y_j(t)$ "lag behind" in $t$ in the $n^* = 6n_0$ approximation by $\approx 20\%$ as compared with the less realistic $n^* \gg n_0$ approximation. When the equilibrium distribution $\{n_j(\infty)\}$ is achieved, the values $y_j$ in the $n^* = 6n_0$ approximation exceed those in $n^* \gg n_0$ approximation, i.e. space charge in the equilibrium state is more "diffuse" in the $n^* = 6n_0$ approximation; (iii) at the EDL relaxation, the transition of the basic charge occurs from the $X^{21}$ plane (most distant from the EC surface) to the $X^1$ plane accompanied by a decrease of the heterojunction voltage. It follows that EDL-capacitance ($C_{EDL}$) rises at $t \to \infty$ and $\sum n_j = const$. It agrees with the experimental data presented in the reviews [50,51].

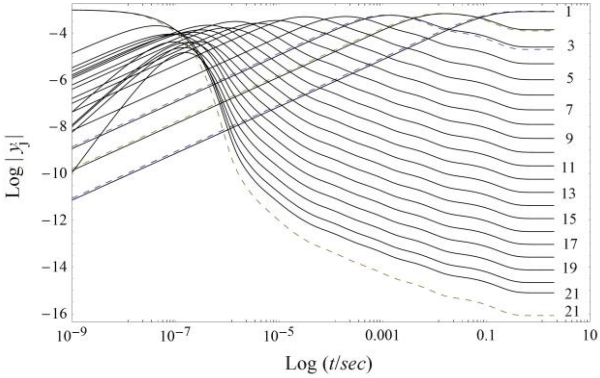

Fig.4. Time-dependence of the moduli of relative changes in the cation concentrations $|y_j| \equiv |n_j - n_0|/n_0$ in log-log scale during space charge relaxation on the EC/$\{X^j\}$ ($j = 1,2...21$) heterojunction. The calculation parameters are $n_0 = 10^{18}$ m$^{-2}$ and $T = 300$ K. The graph numbers from 1 to 21 correspond to the $j$ index of the $X^j$ crystallographic plane. Solid lines of graphs correspond to the $n^* = 6n_0$, and dashed lines to the $n^* \gg n_0$ approximations.

### 3.2. *Effective thickness of the electric double layer*

The effective thickness $\lambda_{eff}$ of EDL on EC/SE-AdSIC blocking heterojunctions can be determined as the thickness of an ionic space charge layer which significantly, e.g. by 5-10 times, decreases the electric field of the EC surface charge. The value $\lambda_{eff}$ is connected with EDL-capacitance $C_{EDL} \propto 1/\lambda_{eff}$. The analysis of the data (Fig. 4) shows that capacitance $C_{EDL}$ is time dependent. Hence, $C_{EDL}$ should depend on the frequency $\omega/2\pi$ of ac influence on EC/$\{X^j\}$. In the structure-dynamic approach $C_{EDL}$ weakly depends on $\omega/2\pi$ in two cases: (1) at $\omega/2\pi < 1/\max\{\tau_r\}$, where $\max\{\tau_r\}$ is the relaxation time for the highest $\eta_{j,j+1}$ in the irregular potential relief; (2) at $\omega/2\pi \sim 10^6$ -$10^7$ Hz $\ll 1/\tau_v \sim 10^9$ Hz, where $\tau_v$ is the relaxation time of non-equilibrium charge in the AdSIC bulk (small height $\eta$, large coordinate $x$, $\lambda_{eff} \approx M\Delta$. In the intermediate range $1/\max\{\tau_r\} < \omega/2\pi < \sim 10^6$ -$10^7$ Hz DEL-capacitance $C_{EDL}$ decreases, and $\lambda_{eff}$ increases with increasing $\omega/2\pi$. In the work [52], the estimation $\lambda_{eff} \approx 1$ nm was used for an equilibrium EDL ($\omega/2\pi \to 0$). This value is based on the "frequency-independent" EDL-capacitance $C_{EDL} \sim 10$ μFcm$^{-2}$ (300 K). The model of adsorption relaxation of EDL [50,51] connects the value $\sim 10$ μFcm$^{-2}$ with mobile ions, and additional large adsorption capacitance $C \gg 10$ μFcm$^{-2}$ and Warburg impedance - with minority charge carriers.

The structure-dynamic approach relates the "frequency-independent" capacitance $\sim 10$ μFcm$^{-2}$ to the frequency range $1/\max\{\tau_r\} \ll \omega/2\pi \sim 10^6 - 10^7$ Hz $\ll 1/\tau_v \sim 10^9$ Hz, for which $\lambda_{eff} \approx M\Delta$. After the termination of relaxation processes in EDL at $\omega/2\pi \ll 1/\max\{\tau_r\}$, our calculation gives $\lambda_{eff} \approx 0.1$ nm ($n_0 = 10^{18}$ m$^{-2}$ and $T = 300$ K) for both $n^* = 6n_0$ and $n^* \gg n_0$ approximations. The equilibrium concentrations $\{y_j(\infty)\}$ ($j = 1, 2, 3, 4$ and 5) for both $n^* = 6n_0$ and $n^* \gg n_0$ approximations can be estimated by the dependence $y_j = \beta \exp(-j\Delta/\lambda_\infty)$, where $\beta \propto n_0/T$ with an accuracy $\approx 20$ % and $\lambda_\infty \propto (T/n_0)^{1/2}$ accurate to $\approx 30$ % for $n_0$ in the range $4 \cdot 10^{17}$ - $10^{18}$ m$^{-2}$ and $T$ in the range 280 - 380 K. The proposed structure-dynamic approach describes the fast and slow ion-transport processes in the uniform way than cardinally differs from the model [50,51] of adsorption relaxation of the EDL. Fig. 4 shows that the main fraction of space ionic charge in EDL does not extend further than the first two $X^j$ planes in equilibrium state. This agrees with the old conclusions made by Remez and Chebotin (1984) [53], that " *…the Debye radius is much smaller than interior spacing …* " and "*… main contribution to the double layer capacitance is made by the charging of the first two monolayers …*" in superionic conductors.

### 3.3. *Ionic transport at ac external influence*

Ionic net currents $I_{j \to j+1}$ for highest $\eta_{j,j+1}$ are shown in Fig. 5 for the case of ac external influence ($\omega/2\pi = 4$ Hz) on the EC/$\{X^j\}$. The $\eta_{j,j+1}$ heights are determined by (1). Current density $I_{21}(t) = I_{21}\sin\omega t$ is created on the $X^{21}$ edge plane by current generator. The amplitude $I_{21}$ satisfies the criterion (18) of smallness $I_{21} << |I_{j\to j+1}|$ for any barrier $\eta_{j,j+1}$. Calculations show that the ionic currents $I_{j,j+1}$ over the highest barriers $\eta_{j,j+1}$ begin to decrease appreciably with increasing $\omega/2\pi$ and lag in phase from the external current $I_{21}(t)$. It corresponds to the dielectric behavior. The calculations for both $n^* = 6n_0$ and $n^* >> n_0$ approximations show that the influence of an occupancy degree of the nearest neighboring positions on the $I_{j,j+1}(t)$ values is small in comparison with such factors as $\omega/2\pi$ and barrier height $\eta_{j,j+1}$. The potential difference between the $X^j$ plane (right edge of the EC/$\{X^j\}$ ($j = 1,2,\ldots M$) heterojunction) and EC can be determined as

$$U_{M,0}(t) = \sum U_{j+1,j}(t), \quad 0 \le j \le M - 1 \quad (19)$$

where $U_{j+1,j}(t)$ designates a potential difference between the $X^{j+1}$ and $X^j$ planes. For example,

$$U_{6,5}(t) = (e\Delta/\varepsilon_0 \varepsilon_{5,6}) \sum (n_j - n_0), \quad 6 \le j \le M \quad (20)$$

where $\varepsilon_{5,6}$ is the effective relative dielectric permeability of the layer between $X^5$ and $X^6$ planes. The absence of electrochemical reaction on the EC blocking electrode can be presented as $U_{M,0}(t) \le 0$ for AdSICs with cation conductivity. The heterojunction voltage can be determined as $V_{M,0}(t) \equiv -U_{M,0}(t)$.

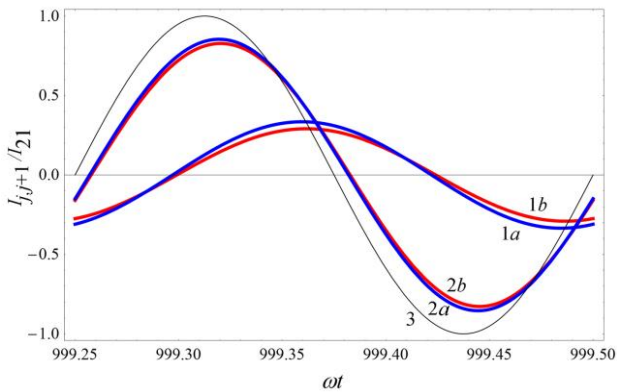

Fig.5. Dimensionless currents $I_{12}(t)/I_{21}$ and $I_{23}(t)/I_{21}$ for highest $\eta_{1,2} = 0.7$ eV and $\eta_{2,3} = 0.62$ eV on the EC/$\{X^j\}$ ($j = 1,2,\ldots 21$) and $I_{21}(t)/I_{21}$ created by the current generator on the $X^{21}$ at the excitation $I_{21}(t) = I_{21}\sin\omega t$. The calculation parameters are $\omega/2\pi = 4$ Hz, $T = 300$ K, $n_0 = 10^{18}$ m$^{-2}$. $1a$ corresponds to $I_{12}(t)/I_{21}$ ($n^* >> n_0$ approximation); $1b$ corresponds to $I_{12}(t)/I_{21}$ ($n^* = 6n_0$); $2a$ corresponds to $I_{23}(t)/I_{21}$ ($n^* >> n_0$); $2b$ corresponds to $I_{23}(t)/I_{21}$ ($n^* = 6n_0$); $3$ corresponds to $I_{21}(t)/I_{21}$.

As was noted in Introduction, an opportunity to use the diffusion-drift equations with constant coefficients $D$ and $\mu$, which are proportional to ionic conductivity $\sigma$ since $D = \mu k_B T e^{-1}$ in the linear approach, should be specially substantiated in nanoionics. To ascertain the criterion for the diffusion-drift equations application, let consider the behavior of $\sigma_{j,j+1}$ in the frame of the structure-dynamic approach

$$\sigma_{j,j+1} \equiv I_{j,j+1} \Delta / V_{j+1,j} \quad (21)$$

where relations (12) and (19) determine the voltages $V_{j+1,j}(t)$ and ionic currents densities $I_{j,j+1}(t)$ in the EDL region. If the quantities $\sigma_{j,j+1}$ do not depend on $t$, $\omega/2\pi$ and current amplitude $I_{21}$ of external influence on a heterojunction, then quantities $\sigma_{j,j+1}$ correspond to the definition of specific ionic conductivity (Ohm law).

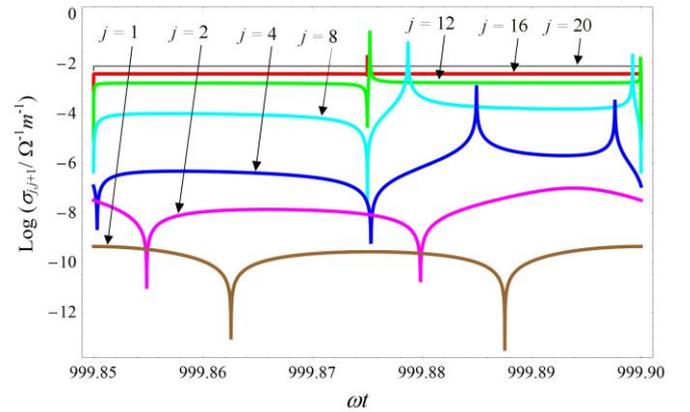

Fig.6. Dependence of logarithms $\sigma_{j,j+1}$ on the phase $\omega t$ at external influence on the EC/$\{X^j\}$ ($j = 1,2,\ldots 21$) at $I_{21}(t) = I_{21}\sin\omega t$ for $\omega/2\pi = 0.1$ Hz. The behavior of $\sigma_{j,j+1}$ at $j > 12$ corresponds to the definition of specific ionic conductivity.

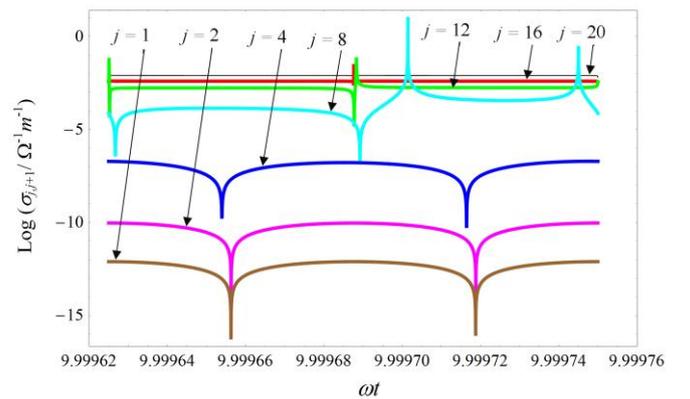

Fig.7. Dependence of logarithms $\sigma_{j,j+1}$ on the phase $\omega t$ at external influence on the EC/$\{X^j\}$ ($j = 1,2,\ldots 21$) at at $I_{21}(t) = I_{21}\sin\omega t$ for $\omega/2\pi = 20000$ Hz. The behavior of $\sigma_{j,j+1}$ at $j > 12$ corresponds to the definition of specific ionic conductivity.

The data presented in Fig. 6 and Fig. 7 show that quantities $\sigma_{j,j+1}$ at $j > 12$ do not practically depend on $t$ in the frequency range 0.1 - 20000 Hz, i.e. the Ohm law holds for average space variation $\eta$ less than $\approx$ 0.04 eV/nm as it follows from equation (1). Thus, we show that the criterion for application of transport equations with constant $D$ and $\mu$ is smallness of $\eta$ variations on nanoscale.

### 3.4. *Impedance of heterojunction and hidden variables*

The method of complex amplitudes is frequently used in the linear approximation of experimental data in impedance spectroscopy. For EC/{$X^j$}heterojunctions the method consists in the replacement of the real variables by complex ones

$$I_M(t) \to I \exp(\mathbf{i}\omega t) \quad (22)$$

$$V_{M,0}(t) \to V_{M,0} \exp(\mathbf{i}\omega t + \mathbf{i}\varphi) \quad (23)$$

The impedance $Z$ of EC/{$X^j$}heterojunction can be defined as

$$Z \equiv V_{M,0} \exp(\mathbf{i}\omega t + \mathbf{i}\varphi) / I \exp(\mathbf{i}\omega t) \quad (24)$$

where the real quantities $V_{M,0}$ and $I$ are the amplitudes of the heterojunction voltage and current density on the $X^M$ plane (current induced by the current generator), $\mathbf{i}$ is the imaginary unit, and the phase $\varphi < 0$ at the capacitive behavior of $Z$. Calculations in the frame of the structure-dynamic approach are executed in the conditions where the approximation of data on the frequency behavior $Z$ could be considered linear (an error less than 0.1 % at $I$ changes by 100 - 1000 times).

For the EC/{$X^j$}blocking heterojunctions (Fig.8), a general tendency of $Z$ behavior consists in that Re- and Im-components of $Z$ decrease with an increase of $\omega/2\pi$, $n_0$ and $T$. The calculations of $Z$ in the $n^* = 6\,n_0$, and $n^* \gg n_0$ approximations show that the degree of occupancy of the potential relief positions accessible for mobile ions does not significantly affect the form of the Nyquist plots. The impedance $Z$ of the EC/{$X^j$}blocking heterojunction is the result of averaging over a set of internal voltages $V_{j+1,j}$ and currents $I_{j,j+1}$, i.e. ion jumps $j \leftrightarrow j \pm 1$, coordinated by the long-range Coulomb interaction.

Figure 8 (Nyquist plot) presents the data on modeling the frequency behavior of impedance $Z$ of the EC/{$X^j$}with the profile of potential relief (1). Figure 9 presents the dependence of the energy fraction $\int I_{j,j+1} V_{j+1,j} dt \;/\; \int I_M(t) V_{M,0}(t) dt$ (hidden physical quantities) dissipated within the period $2\pi/\omega$ of external influence on the layer between $X^j$ and $X^{j+1}$ planes in EDL.

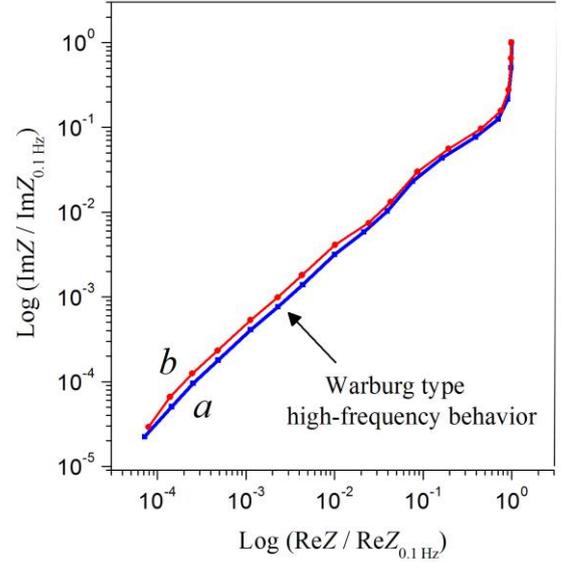

Fig. 8. Nyquist plot for Re- and Im-components of impedance $Z$ of the EC/{$X^j$} ($j = 1,2,.. 21$) blocking heterojunction with the potential relief profile described by (1). The calculations are made for the $n^* = 6n_0$ (graph *a*), and $n^* \gg n_0$ (graph *b*) approximations in the frequency range of 0.1 - 20000 Hz. The calculation parameters are $n_0 = 10^{18}$ m$^{-2}$, $T = 300$ K.

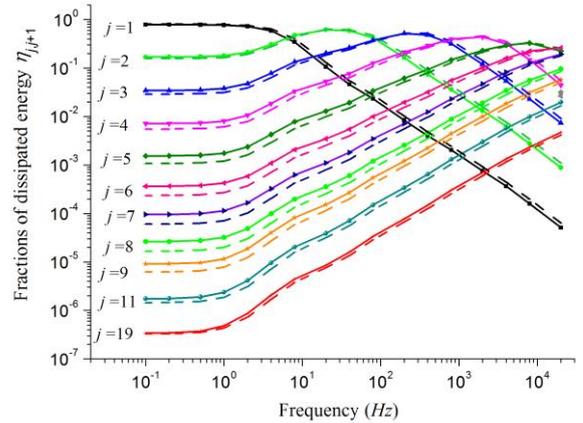

Fig.9. The fractions of dissipated energy on separate layers of EDL on the EC/{$X^j$} ($j = 1,2,.. 21$) blocking heterojunction (between $X^j$ and $X^{j+1}$ planes) with the potential relief profile described by equation (1). The numbers of graphs correspond to the $j$ index of the potential barrier $\eta_{j,j+1}$. The solid lines correspond to the $n^* = 6n_0$ approximation, and dotted lines correspond to $n^* \gg n_0$. The calculation parameters are $n_0 = 10^{18}$ m$^{-2}$, $T = 300$ K.

It follows from the comparison of Fig. 8 and Fig. 9 that a set of hidden quantities, e.g. integrals $\int I_{j,j+1} V_{j+1,j} dt$, contains much more information about processes in EDL than Re$Z$- and Im$Z$-impedance components which are accessible to direct measurement in experiment.

The section of the "near constant slope" on the Nyquist plot (Fig. 8) corresponds to the frequency range of Fig. 9 where the energy dissipation decreases with increasing $\omega/2\pi$. The analysis of frequency dependence of ionic currents $I_{j,j+1}$ shows that the concentration of cations incapable to jump over high potential barriers $\eta_{j,j+1}$ ($j = 1,2…$) during the half-period $\pi/\omega$ of external influence rapidly grows at $2\pi/\omega$ increasing (dielectric behavior). At further $2\pi/\omega$ increase the dielectric behavior extends to lower potential barriers $\eta_{j,j+1}$ in the EDL-region.

The lines in the graphs in Fig. 8 have a slop of approximately $\pi/4$ in the frequency range $1/\max\{\tau_r\} \ll \omega/2\pi < 1/\tau_v$. It is known [54] that a slope of $\pi/4$ is a characteristic of the Warburg impedance ($W$) with $\mathrm{Re}W$ and $\mathrm{Im}W \propto (\omega D)^{0.5}$, where $D$ is the diffusion coefficient. In the case of a potential relief with a profile described by (1), diffusion and drift cannot be characterized by definite values of the coefficients $D$ and $\mu$ if $\omega/2\pi \sim 1/\max\{\tau_r\}$. An increase of the external influence frequency to $\omega/2\pi \gg 1/\max\{\tau_r\}$ creates conditions where most of mobile cations do not succeed in jumping over high potential barriers during the half-period $\pi/\omega$. Therefore the ionic transport over lower barriers with $j > 10$ and the barrier heights profile close to $\eta_{j,j+1} \approx const$ mainly contributes to the Re-component of heterojunction impedance. A chain of such barriers on the distance $l$ that considerably exceeds the length of an ion elementary jump $l \sim \Delta M/2 \gg \Delta$. It corresponds to the condition of diffusion in a regular potential relief. Thus, in the frequency range $1/\max\{\tau_r\} \ll \omega/2\pi < 1/\tau_v$ the impedance of the EC/$\{X^j\}$ ($j = 1,2…21$) blocking heterojunction with a profile described by (1) can be expressed through the value of diffusion coefficient $D$, like in the case of the Warburg impedance.

### 3.5. *Advanced superionic conductors: ionic transport in galvanostatic mode*

Ionic transport was modeled for galvanostatic charging of the EC blocking electrode contacting with AdSIC. The distribution of potential barrier heights in an EC/$\{X^j\}$ ($j = 1,2…21$) blocking heterojunction is determined by the dependence

$$\eta_{j,j+1} = a_1 + b_1\exp(-0.5j/c), \text{ b}(j = 1,2…20) \quad (25)$$

where $a_1 = 0.1$ eV, $b_1 = 0.55$ eV and $c = 4.5$, i.e. in the AdSIC bulk the potential barriers height is $\approx 0.1$ eV. Dependence (25) corresponds to crystallographically narrow interfaces. The specified difference of potential barrier heights $0.5 - 0.1$ eV corresponds to the transition from cation-conducting SE with a rather low ionic conductivity to AdSIC, e.g. $\alpha$-RbAg$_4$I$_5$ family with ionic conductivity $\sigma > 0.2$ Ohm$^{-1}$ cm$^{-1}$ (300 K) [17].

In the galvanostatic mode a current generator creates a specified density of ionic current according to the law: $I_{21}(t) = 0$ at $t \leq 0$ and $I_{21}(t) = const \neq 0$ at $t > 0$ on the $X^{21}$ edge plane. The ratio of surface charge density $Q(t)$ accumulated on EC/$\{X^j\}$ ($j = 1,2…21$) to voltage $V_{M,0}(t)$ on a blocking heterojunction, i.e. to the voltage between EC and $X^M$, can be considered as capacitance of the heterojunction. Calculations in the approximation $n^* = 6n_0$ show (Fig. 10) that for galvanostatic charging the capacitance $Q(t)/V_{M,0}(t)$ increases with $t$ and reaches large (small) values in the limit $t \to \infty$ ($t \to 0$). The capacitance $Q(t)/V_{M,0}(t)$ can be measured in experiment, but the set of hidden variables $\{n_j\}$ determining the behavior of $Q(t)/V_{M,0}(t)$, can be found only by computer modeling.

Steric effects [55,56] (for example, reduction of $C_{EDL}$ at charging of an EC/$\{X^j\}$ ($j = 1,2…21$) heterojunction at galvanostatic charging) become appreciable if $Q(t)$ increases to

$$Q(t) = \int I_{21}(t)\, \mathrm{d}t \sim 0.01 en_0 \quad (26)$$

when $C_{EDL}$ behavior is defined not only by the reduction of cation concentration $n_1$ on the $X^1$ plane, but also by $n_2$ on the $X^2$ plane. Fig. 10 presents the data for the case when the steric effect is still small (~ 0.1 % of $C_{EDL}$ in the maximum) at $t \sim 0.1$ s. A generalized description of changes of charge distribution in the EDL-region is given by the function

$$\lambda(t) = \Delta \sum j\, y_j(t) / \sum y_j(t) \quad (27)$$

with sum over $j$ ($j = 1,2,..\ 21$). The quantity $\lambda(t)$ shows (in the units of $\Delta j$ product) the position of $x$-coordinate of the center of mass for cation deficiency in EDL (from the plane $X^0$, i.e. from EC surface) in the moment $t$. At increasing $t$ the center of cation deficiency charge moves from the $X^{21}$ plane to the $X^1$ plane.

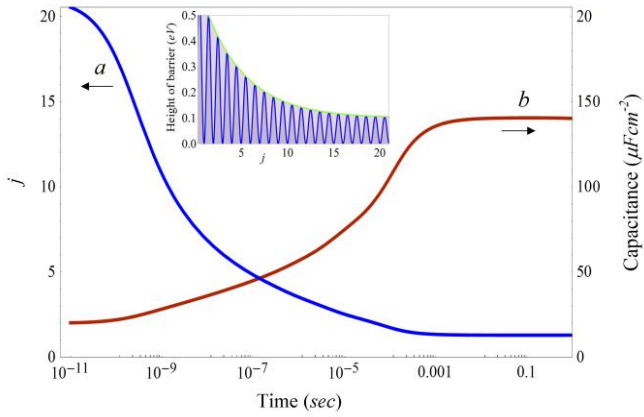

Fig.10. Galvanostatic charging of the EC/{$X^j$} ($j$= 1,2,.. 21) blocking heterojunction. Time dependent functions of the $\lambda(t)$ center of mass for cation distribution (*a*) and of the $Q(t)/V_{M,0}(t)$ capacitance density (*b*) for a sequence of potential barriers (insert) determined by (25).

Time scales at $\lambda(t)$ change are defined by the distribution of potential barrier heights $\eta_{j,j+1}$ in the irregular potential relief with a profile determined by (25). Thus, the heterojunction capacitance considerably increases during relaxation, which corresponds to the experimental data presented in [11,40].

## 4. Conclusions

- In nanoionics, the problem of searching of simple physico-mathematical approach giving: (1) the detailed description of ion-transport processes (direct problem) in an irregular potential relief of a solid electrolyte (SE), and (2) the interpretations of properties and characteristics of nanosystems with fast ion transport (inverse problem) was formulated.

- The proposed structure-dynamic approach includes: (i) a structural model that interconnects the relaxation rate of an electric double layer (EDL) and ion movement in the crystal potential relief of a "rigid" sub-lattice of SE distorted at the SE/electronic conductor (EC) heterojunction; (ii) method of "hidden" variables providing the description of the ion-transport processes and solution of the inverse problems in terms of mobile ion concentrations on the crystallographic planes of SE in the region of space charge; (iii) physico-mathematical formalism operating with "hidden" variables and based on the concept of a detailed balance and a kinetic equation in the form of the particle conservation law.

- A computer simulation of ionic transport in the region of space charge on an ideally polarizable SE/EC heterojunctions is performed and the frequency (time) behavior of SE/EC-capacitance and impedance are calculated for the external influence of current generator (ac and galvanostatic modes).

- It is proposed to characterize the distribution of space charge in the EDL by position of center of mass of ion deficiency.

- The data about position changes of the center of mass of a space charge during the EDL relaxation are obtained for galvanostatic mode.

- The criterion for application of transport equations with constant diffusion and drift coefficients is smallness on nanoscale the variations of potential relief depth.

- The proposed structure-dynamic approach in nanoionics describes the fast and slow ion-transport processes in the uniform way than cardinally differs from known model of adsorption relaxation of the EDL.

- The structure-dynamic approach can find application in the development of new nanodevices with FIT, e.g. supercapacitors with structure-ordered heterojunctions necessary for the development of deep-sub-voltage nanoelectronics [17] and microsystem technology.

## Acknowledgments

We thank Prof. Y.M. Gufan (Rostov-on-Don State University, Russia) proposed (2012) an exponential approximation for profile of potential relief.